\documentclass[preprint,review,12pt]{elsarticle}

\usepackage{graphicx}
\def\be{\begin{eqnarray}}
\def\ee{\end{eqnarray}}
\def\hfg{{\tt HFgrad}}
\def\Tr{{\rm Tr}}
\usepackage{amssymb}

\newcounter{bla}

\journal{Computer Physics Communications}

\begin{document}

\begin{frontmatter}


%Title\tnoteref{label1}}
\author{G.F.~Bertsch} 
\author{J.M.~Mehlhaff}
\address{Institute for Nuclear Theory and Dept. of Physics, Box 351560,
University
of Washington, Seattle, Washington 98915, USA}

%  \address{Address\fnref{label3}}

%\title{A \LaTeX{} template for CPC Computer Physics Descriptions}
\title{A finite-temperature Hartree-Fock code for shell-model
Hamiltonians} 

\cortext[a] {Corresponding author.\\\textit{E-mail address:} 
bertsch@uw.edu}
%\address[a]{First Address}
%\address[b]{Second Address}

\begin{abstract}
The codes {\tt HFgradZ.py} and {\tt HFgradT.py} find axially 
symmetric minima of a
Hartree-Fock energy functional for a Hamiltonian supplied in a shell
model basis.  The functional to be minimized is the 
Hartree-Fock energy for zero-temperature properties
or the Hartree-Fock grand potential for finite-temperature 
properties (thermal energy, entropy).
The minimization may be subjected to additional constraints besides the
neutron and proton numbers.  A single-particle operator can be used
to constrain the minimization by adding it to the single-particle
Hamiltonian with a Lagrange multiplier. One can also constrain its 
expectation value in the zero-temperature code.  Also the orbital 
filling can be constrained in the zero-temperature code,
fixing the number of nucleons having given $K^\pi$ quantum numbers.  
This is particularly useful to resolve near-degeneracies among distinct minima. 
\end{abstract}


\end{frontmatter}

% \linenumbers

% Computer program descriptions should contain the following
% PROGRAM SUMMARY.

{\bf PROGRAM SUMMARY}
  %Delete as appropriate.

\begin{small}
\noindent
{\em Manuscript Title:}  
A finite-temperature Hartree-Fock code for shell-model
Hamiltonians  \\
{\em Authors:}   G.F. Bertsch and J.M. Mehlhaff            \\
{\em Program Title:}  HFgradZ.py, HFgradT.py                          \\
{\em Journal Reference:}                                      \\
  %Leave blank, supplied by Elsevier.
{\em Catalogue identifier:}                                   \\
  %Leave blank, supplied by Elsevier.
{\em Licensing provisions:}                  none             \\
  %enter "none" if CPC non-profit use license is sufficient.
{\em Programming language:}   Python (2.7)              \\
{\em Computer:}  PCs                                              \\
  %Computer(s) for which program has been designed.
{\em Operating system:} Unix, Apple OSX                       \\
  %Operating system(s) for which program has been designed.
{\em RAM:} 10 MBy                                              \\
  %RAM in bytes required to execute program with typical data.
  % Please give some freely chosen keywords that we can use in a
  % cumulative keyword index.
{\em Classification: 4.9 Minimization and fitting, 17.22 Hartree-Fock}                                         \\
  %Classify using CPC Program Library Subject Index, see (
  % http://cpc.cs.qub.ac.uk/subjectIndex/SUBJECT_index.html)
  %e.g. 4.4 Feynman diagrams, 5 Computer Algebra.
{\em External routines/libraries:} Numpy  (1.6)                \\
  % Fill in if necessary, otherwise leave out.
  %Only required for a New Version summary, otherwise leave out.
{\em Nature of problem:}\\  
Find Hartree-Fock minima of shell-model Hamiltonians
   \\
{\em Solution method:}\\  Gradient method with a preconditioner\\
  %Describe the method solution here.
{\em Running time:}\\  a few minutes 
  %Give an indication of the typical running time here.
   \\

\end{small}

\section{Introduction}
  The usual methods for finding the Hartree-Fock (HF) minima of nuclear 
Hamiltonians focus on the equations
that must be satisfied at the minimum,
\be
{ d \langle H \rangle \over d \vec x} = 0
\ee
Here $\langle H \rangle$ is the Hartree-Fock expression for the energy,
Eq. (\ref{EHF}) below, and $\vec x$ is the set of variational parameters.
Convergence problems can easily arise, as documented in Sect. 5.4 of Ref.
\cite{RS}.
They may be overcome by sophisticated iteration schemes such as the Broyden
method \cite{ba08}, but we find that the gradient method advocated
in Ref. \cite{RS} and adopted  Refs. \cite{robledo,
ro11} is simple and effective.  The gradient method is implemented
in \hfg~by constructing the vector $d \langle H \rangle / d\vec x$
and using it to guide the iteration process.  This is described
in Sect. 3 below.

\section{Variables}
We consider a basis of shell-model wave functions labeled by
$\ell,j,m$ and $\tau_z$ and distinguished by an index $i,j,...$.
The orbitals are linear combinations of the shell-model wave
functions; they are indexed by Greek letters $\kappa,\lambda,...$.
The many-body wave function is characterized by an orthogonal matrix
$U$ that transforms from the shell basis to the orbital basis
\be
|\kappa\rangle = \sum_i U_{\kappa,i} |i\rangle
\ee
and a diagonal matrix $P$ specifying the occupation factors in
the orbital basis
\be
\label{P}
P_{\kappa\lambda}  = \delta_{\kappa\lambda} f_\kappa.
\ee
In the zero-temperature code, 
$f_\kappa = 1$ or 0 depending on whether the orbital is
occupied or not, while it the finite temperature code it can
vary between these limits. 

The fundamental physical quantity associated with the HF solution
is the single-particle density matrix $\rho$, given by
\be
\label{rho}
\rho = U^T P U.
\ee
The nominal dimension of the matrix is $N_d= N_b^2$, where $N_b$ is the
number of states in the shell-model basis.  However, the restriction in
the code to axially symmetric configurations with good parity considerably
reduces the number of nonzero terms; the code takes advantage of the
symmetry by separating the matrix into blocks.

\section{Basic equations}
The code treats Hamiltonians that can be represented as a sum of a 
diagonal one-body
operator in Fock space together with a two-body interaction given by
its $J$-coupled matrix elements.  The basis states are the shell-model
states $|i\rangle = | \tau_{zi}, \pi_i, j_i,m_i\rangle$ where $\tau_z$ is the isospin, $j$ is
the angular momentum, 
$\pi$ is the parity and $m$ is the $z$-component of angular
momentum.  The input Hamiltonian may be written
\be
H = {\hat K} + {\hat v}
\ee
where
\be
{\hat K} = \sum_i \varepsilon_i {\hat a}^\dagger_i {\hat a}_i 
\ee
\be
{\hat v} = \sum_{i<j,k<l} v_{ij,kl}\, {\hat a}^\dagger_i {\hat a}^\dagger_j 
{\hat a}_l {\hat a}_k 
\ee
and
\be
\label{vijkl}
v_{ij,kl} = d_{ij} d_{kl}\sum_{JM}(i\,j|v|k\,l)_J\, 
(j_i \,j_j\, m_i\, m_j|J \,M)(j_k \,j_l \,m_k \,m_l|J\, M)
\ee
Here $(i\, j|v|k\,l)_J$ are the $J$-coupled interaction matrix
elements, $(j_i \,j_j \,m_i \,m_j|J\, M)$ are Clebsch-Gordon recoupling
coefficients, and
$d_{ij} = (1+ \delta_{ij})^{-1/2}$.
Only the elements $\rho_{ij}$  with $(\tau_{zi},\pi_i,m_i)=(\tau_{zj},\pi_j,m_j)$
are kept in the array representing the density matrix; the imposed
symmetries require that other elements are zero.  
Similarly, only terms that can give nonzero contributions to 
the interaction energy are kept in the array representing ${\hat v}$.

Both the energy and the gradient are compute using the
single-particle potential $V$ as an intermediate array.  It is 
defined
\be
\label{V}
V_{ik} = \sum_{ijkl} (v_{ij,kl}-v_{ij,lk}) \rho_{jl}.
\ee
The gradient is derived from
the single-particle Hamiltonian
\be
\label{Hsp}
H^{sp} = K + V,
\ee
a matrix with nominal dimension $N_d \times N_d$.  

Besides $\rho$, the matrices $U,\rho,V$ and $H$ are
block-diagonal with the blocks determined by $(\tau_z,\pi,m)$.  The code
takes advantage of the structure to store these matrices in packed arrays.
The two-body interaction is also stored in a packed array that allows
Eq. (\ref{V}) to be evaluated by ordinary matrix-vector multiplication.

For convenience the code is split into two driver
modules,  {\tt HfgradZ.py} for HF at zero temperature and 
{\tt HFgradT.py}
for finite temperature. 
The zero-temperature code minimizes the HF energy
\be
\label{EHF}
E = \langle H \rangle = \Tr K\rho + {1\over 2} \Tr_2 \rho v \rho.
\ee
The finite-temperature code minimizes the grand potential.
In terms of $\rho$ and $f_\kappa$, the
grand potential at inverse temperature $\beta$ is given by
\be
\label{omega}
\Omega = E - \beta^{-1} S  +\left( \sum_{\tau_z} \mu_{\tau_z}
N_{\tau_z}\right)
\ee
with $E$ from Eq. (\ref{EHF}), entropy $S$ given by
\be
\label{S}
S = \sum_\kappa \left(f_\kappa \ln f_\kappa +
(1-f_\kappa) \ln( 1 - f_\kappa)\right),
\ee
and the expectation values of particle number $N_{\tau_z}$ in
the last term.  The latter are segregated in parentheses because
that term has no role in the gradient evaluation;
the minimization will be carried out at fixed 
$N_{\tau_z}$.

\section{The hybrid minimization method}

The minimization with respect to the elements of $U$ is same in 
both codes.  
The constraint that $U$ is orthogonal is satisfied in the iterative process
by starting with an orthogonal matrix and updating it by an explicitly
orthogonal transformation.  
The update from $U$ to $U'$ can be expressed as
a Thouless transformation of $U$,
\be
\label{Z}
U' = e^Z U.
\ee
Here $Z$ is a skew-symmetric matrix of the independent
variables $z_{\kappa\lambda}$ ($\kappa < \lambda$), giving 
$N_d(N_d-1)/2$  variational parameters in the general case, i.e. without any 
conserved quantum numbers.  The
block structure associated with the $(\tau_{zi},\pi_i,m_i)$ quantum numbers 
greatly reduces that number. 

The gradient of $E$ (Eq. (\ref{EHF}) with respect to the elements of the 
$Z$ matrix is performed analytically to arrive at the expression
\be
{\partial E \over \partial z_{\kappa\lambda}} =
H^{orb}_{\kappa\lambda}(f_\kappa - f_\lambda).
\ee
Here $H^{orb}$ is the single-particle Hamiltonian in
the orbital basis,
\be
\label{Horb}
H^{orb}  = U H^{sp} U^T.
\ee
Given the gradient, the simplest algorithm to update $U$ is the steepest
descent method.  Here one would use Eq. (\ref{Z}) with
\be
\label{eta}
z_{\kappa\lambda} = 
\eta_z {\partial E \over \partial z_{\kappa\lambda}}.
\ee
where $\eta_z$ is some small numerical parameter that controls the
stability of the algorithm and its convergence rate. However, 
convergence of the steepest descent iteration is often poor.
A much more efficient algorithm is used by Robledo in
his HFB code \cite{robledo}.  It takes into account approximately the curvature
of the
energy surface by introducing a preconditioner into right-hand side of Eq.
(\ref{eta}).  

The present code employs a different method that achieves the same
purpose, which we call the hybrid method.  At each iteration
step, the code diagonalizes a 
modified orbital Hamiltonian $H^{orb}_\eta$ with the same diagonal
elements as $H^{orb}$ but reduced off-diagonal elements:
\be
\label{hybrid}
H^{orb}_{\eta}|_{ij} = \delta_{ij} H^{orb}_{ii}+ \eta_z
(1-\delta_{ij}) H^{orb}_{ij}.
\ee
The transformation matrix  $U_\eta$ that diagonalizes $H^{orb}_\eta$ is
used to update $U$,
\be
\label{update}
U' = U_\eta U
\ee
In the limit 
$\eta_z \ll 1$ the method amounts to a perturbative approximation to the
$U_\eta$,  equivalent to Robledo's 
preconditioned form 
\be
z_{\kappa\lambda} = 
\eta_z {1 \over \left|H^{orb}_{\kappa\kappa}-H^{orb}_{\lambda\lambda}\right|}
{\partial E \over \partial z_{\kappa\lambda}}.
\ee
One caveat: the $U_\eta$ must keep orbitals ordered 
by the diagonal elements $H^{orb}_{\kappa\kappa}$.
The hybrid method also transforms the empty and
filled orbitals among themselves, but that does not change $\rho$ or 
affect any HF observables.

In another limit, 
namely $\eta_z= 1$,  the method amounts to a straightforward diagonalization of the
single-particle Hamiltonian.  This is often part of the update process in non-gradient
methods.  Thus, the hybrid method achieves both update techniques under the
control of a single parameter. 

Part of the update may require forcing a
change in the expectation value of a single-particle operator.  For that
purpose, $U$ is updated by a direct approximation to Eq. (\ref{eta}), as
discussed in the next section.

\subsection{Operator constraints}
Typically, there are many local minima of the Hartree-Fock energy
functional.  They will also be present in the grand potential, becoming
weaker as the temperature of the ensemble increase.  It is
important to permit additional constraints on the solutions beyond those
for the number operators, in order to explore the energy surface and locate the
possible minima.  This is facilitated in the code by allowing the user to 
numerically define a single-particle operator $Q$ and constrain its
expectation value or just add it as fixed external field.
As an external field, the user
supplies a Lagrange multiplier $\lambda_q$ and the gradient is
derived from the single-particle Hamiltonian
\be
\label{lamq}
H^{sp}_\lambda = K + V - \lambda_q Q
\ee
The other option, constraining $\langle Q \rangle$ to some value $q$, requires
the gradient updating algorithm to carry out two tasks.  The first is to 
correct the wave function to bring $\langle Q\rangle$ closer to 
its target value.  This step is based on a $Z$ matrix with elements
given by
\be
\label{reset}
z_{\kappa\lambda} = {q-\langle Q \rangle\over \Tr Q^{ph}(Q^{ph})^T} Q^{ph}
\ee
where 
\be
\label{Qph}
Q^{ph}_{\kappa\lambda} = Q^{orb}_{\kappa\lambda} (f_\kappa
-f_\lambda)
\ee 
and $Q^orb$ is the operator in the orbital basis as in Eq. (\ref{Horb}).
The updating matrix must be orthogonal, but need only approximate
the exponential $e^Z$.  The code uses a simple Pad\'{e} 
approximant to preserve the orthogonal character \cite{robledo2}
\be
\label{unitary}
e^Z \approx (1+Z/2)(1-Z/2)^{-1}.
\ee 

In the presence of the constraint, the $U$ update for minimization must also
be modified to project $Z$ to a direction that keeps $\langle Q \rangle$
fixed.  This is carried out by replacing $H^{orb}$ by
\be
\label{Horb'}
H^{orb'} = H^{orb} - {\Tr(H^{orb} Q^{ph}) \over \Tr(Q^{ph}(Q^{ph})^T)}
Q^{ph}.
\ee

\subsection{Special at zero temperature}

At zero temperature, the occupation numbers $f_\kappa$ are zero or 
one for each orbital.  For the input data, the set $\{f\}$ is specified
by the particle number in each block rather than orbital-by-orbital.
The neutron and proton numbers for the nucleus
is determined by the initial $\{f\}$ array,
$N_{\tau_z} = \sum_\kappa f_{\tau_z,\kappa}$.  Any change in $\{f\}$ is 
discontinuous so there can be no gradient method
to effect a change.  The code permits two alternatives to deal with
the situation. The $\{f\}$ can be kept fixed throughout the iteration process.
As will be shown in the examples, this option gives a very good
control to locate nearly degenerate local minima.  The code also permits
updates of the occupations numbers.  In that option, 
in each iteration cycle the code populates the orbitals with the lowest 
single-particle energies. Those determined by the diagonalization of $H^{orb}$ or
its constrained forms $H^{orb}_\lambda$ and $H^{orb'}$.  

\subsection{Finite temperature }

The finite-temperature code minimizes the grand potential $\Omega$
or equivalently the partition function of the grand canonical
ensemble.  The occupation factors are now real numbers satisfying
$0 \le f_\kappa \le 1$.  
Rather than using $f_\kappa$ directly, the code uses variables
$\alpha_\kappa$ 
related to $f$ by
\be
f_\kappa = {1\over 1 + e^{\alpha_\kappa}}.
\ee
The gradient of $\Omega$ with respect to the $\alpha$ variables can be carried
out independently of the gradient with respect to $z$.  The latter has the
same
form as in the zero-temperature minimization,
\be
{\partial \Omega \over \partial z_{\kappa\lambda}}
= {\partial E \over \partial z_{\kappa\lambda}} =
H^{orb}_{\kappa\lambda}\,(f_\kappa - f_\lambda),
\ee
The gradient with respect to $\alpha$ is given by 
\be
\label{grad-alpha}
{\partial \beta\Omega \over \partial \alpha_{\kappa}} =
\left(\alpha_\kappa- \beta H^{orb}_{\kappa,\kappa} \right)  
f_\kappa (1 - f_\kappa).
\ee
In the code, the updated set $\{\alpha_\kappa'\}$ is computed as
\be
\label{alpha'}
\alpha_\kappa' = (1-\eta_\alpha) \alpha_\kappa + \eta_\alpha
\left(\alpha_\kappa- \beta H^{orb}_{\kappa,\kappa} \right)  
+ \alpha_{\tau_z}.
\ee
Here $\eta_\alpha$ is the coefficient of the gradient.  The second
term is proportional to the gradient times the preconditioner
$\left(f_\kappa
(1 - f_\kappa)\right)^{-1}$.  The last
term is an $\tau_z$-dependent constant that can be interpreted
as $\beta$ times the chemical potential.  It is determined from
the equation $N_{\tau_z} = \sum f_\kappa(\alpha'_{\tau_z})$ where
$N_{\tau_z}$ are the proton and number numbers in the data input.
To $\langle Q\rangle$ at the same time in the $\alpha$
update would be more complicated (see Eq. (21) of Ref. \cite{ro11})
and was not implemented in {\tt HFgradT.py}. 

In practice, we have not found any convergence difficult with
respect to the $\alpha$ update taking $\eta_\alpha  = 1$ as in
other iteration schemes.  Still, it is reassuring
to have a gradient method available for
the $f$ variables:  it guarantees that every
cycle of $U$ and $f$ updates lowers the grand 
potential for sufficiently small $\eta_z$ and $\eta_\alpha$.

\section{Running the codes} 

The user must supply files that specify the shell-model space and 
the one-body and two-body matrix elements of the Hamiltonian in
the space.  The files defining the shell-model space and the shell-model
Hamiltonian follow the convention defined in Ref. \cite{brown}.
Note that Hamiltonian interaction matrix elements are input in
the neutron-proton formalism rather than the isospin formalism.

The input data also includes files of the initial occupation 
numbers $\{f\}$ and the initial basis-to-orbital transformation $U$.
For {\tt HfgradZ}, the occupation numbers refer to blocks and the 
size of the array is equal to the number of blocks.  For
the {\tt HFgradT}, the input occupation numbers refer to orbitals
and the size of the array is the dimension of the orbital space.
Note that only the orbitals with positive $m$ are included
in the array; the orbitals with negative $m$ are treated assuming
that the wave function is invariant under time reversal.

In practice, the initial transformation matrix can be quite crude,
as long as it is an orthogonal matrix.  In several of the examples
below, the initial $U$ is taken as the unit matrix.

One last array required by the code is the
matrix of some one-body
field $Q$ such as the quadrupole operator.  Both $U$ and $Q$ inputs
are in the packed-block array format.
 
The command line input file contains 6 or more lines as follows:\\
Line 1.  Name of file defining the shell-model space;\\
Line 2.  Name of the file defining the shell-model Hamiltonian;\\
Line 3.  Name of file giving the initial occupation numbers $f$ of the
single-particle HF orbitals, followed by a flag:  `F' for fixed
occupation numbers, `U' to update occupation numbers;\\
Line 4.  Name of file defining the initial transformation matrix
$u$;\\
Line 5.  Ground-state code:   $\eta_z$, {\tt conv}, {\tt
itermax};  or \\
Line 5'. Finite-temperature code: $\eta_z$, $\eta_\alpha$, {\tt conv},
{\tt itermax}, $Z$, $N$;\\
Line 6.  Name of file defining a single-particle field
$Q$, flag for constraint status (none = `N',Lagrange = `L',
Constrained = `C'), $\lambda_q$ or $\langle Q \rangle$; \\
Line 7+.  $\beta$ (MeV$^{-1}$)   (one or more lines in {\tt HFgradT}).

\section{Output}
The principal outputs of the code, written to the terminal, are
the number of iterations {\tt niter}, the final energy $E$, and
the expectation value of the quadrupole operator $Q$ or other
single-particle operator provided in the input data.  The
finite-temperature code also reports the entropy of the
ensemble, $S$ in Eq. (\ref{S}).

The code also writes to terminal a table of orbital properties.
The columns are:\\
1) index for the orbital;\\
2) index of the block containing the orbital;\\
3) charge of the nucleon (0 or 1);\\
4) $K$ quantum number;\\
5) parity $\pi$: 0 or 1 for even or odd parity respectively;\\
6) occupation number $f$, integer for zero temperature and 
    floating-point for finite temperature;\\
7) single-particle energy.\\

In addition, the code writes the final $U$ matrix and $f$ array
to files {\tt u\_new.dat} and {\tt n\_new.dat}, respectively.
In the zero-temperature code the file has two lines.  The first
line gives the number of occupied orbitals in each block and can be 
used as an input file to {\tt HFgradZ}. .  The
second line give the occupation number for each orbital in the format
needed by {\tt HFgradT}.  Apart from that, the two files are in proper format
to be used as input to rerun the minimization.  If the minimization is
converged, the rerun should only require one
iteration step.

\section{Two examples}
The examples use input Hamiltonians
for $^{162}$Dy and $^{148}$Sm, taken from
Refs. \cite{al08,oz13}.  The shell scripts below illustrate the various
options available when running the codes.
\subsection{$^{162}$Dy}
\noindent
{\tt dy162Z.sh}:
This script runs the zero-temperature code allowing occupation number
changes during the iteration.  The
final energy, $E= -371.78$ agrees with Table II of Ref. \cite{al15}.\\

\noindent
{\tt dy162\_def-sph.sh}:
This script runs the finite temperature code for several
$\beta$ values in the vicinity of the deformed-spherical
phase transition.  The output quadrupole moments
$\langle Q\rangle_\beta$ are shown in 
Fig. 1.  A phase transition at $\beta \approx 0.83$ MeV$^{-1}$ is 
evident.  This is a well-known artifact of mean-field theory and is
absent in more refined treatments \cite{al08,oz13},\\

\noindent
{\tt dy162ZL.sh,dy162ZC.sh,dy162TZ.sh}:\\
These scripts exhibit the use of a constraining field.  The scripts
with an ``L" add the field with a Lagrange multiplier.  The scripts
with a ``C" constrain the expectation value of the field.  The
zero-temperature input parameters have been chosen to show convergence 
to the same state by both methods.  Here the
the converged solution has $E=-370.23$ and $Q= 587.5$.  

\begin{figure}[tb] 
\begin{center} 
\includegraphics[width=\columnwidth]{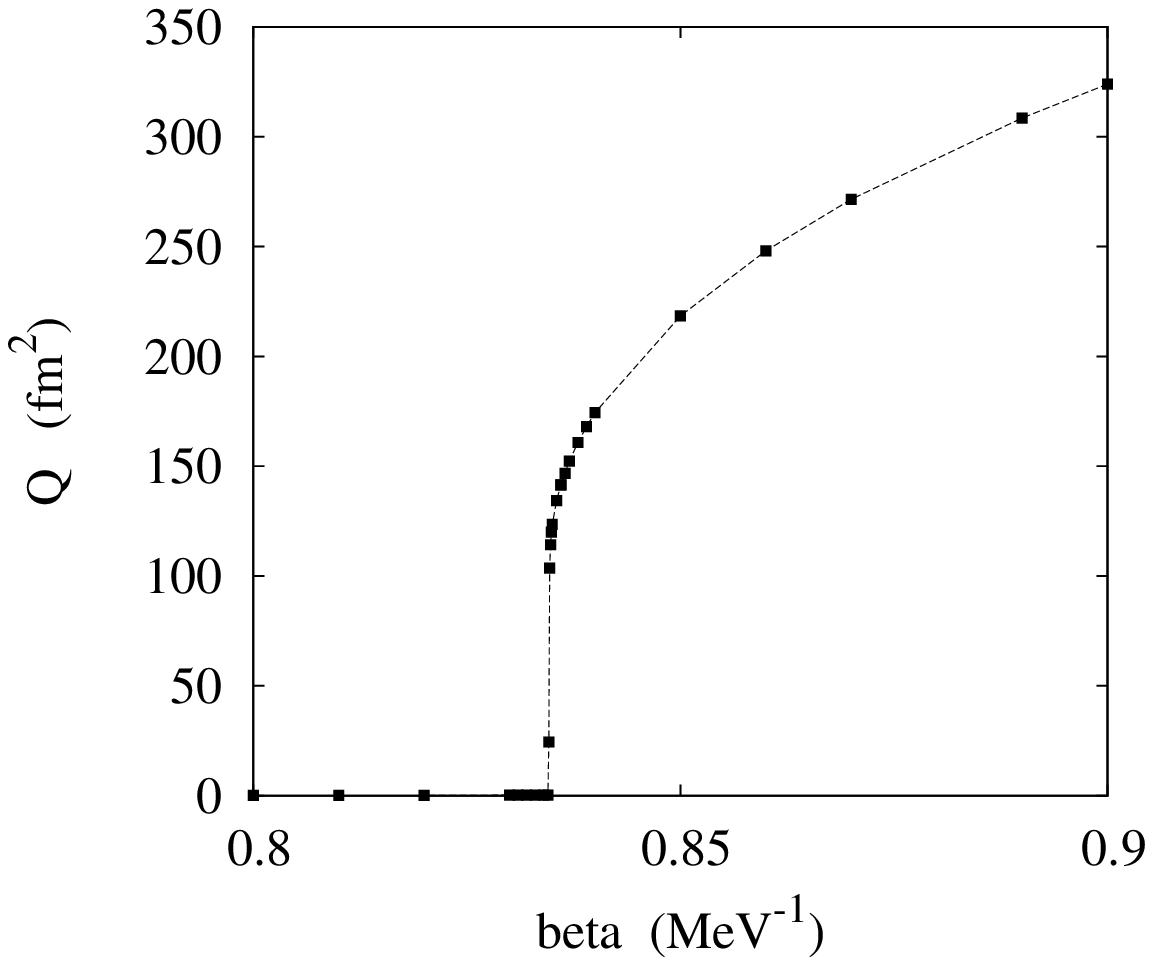} 
\caption{Quadrupole moment as a function of inverse temperature for 
$^{162}$Dy as computed by the {\tt HFgradT.py} code.
}
\label{q_vs_beta} 
\end{center} 
\end{figure} 

\subsubsection{$^{148}$Sm}
\noindent
{\tt sm148U.sh}:
This script shows that the iteration process may fail to converge when
the occupation numbers numbers are allowed to change at each iteration step.
It turns out that the update cycles between 
two sets of occupation numbers.  The two sets differ by a single pair
of neutrons moving between block 
$K^\pi = 1/2^-$ and block $K^\pi = 3/2^-$.\\  
{\tt sm148F.sh}:
This script runs the code for each of the occupation number sets from
the previous script.  There is no longer an oscillation,
and both runs converge.
The total energies of the two
minima are very close to the entry for that nucleus in Table II
of Ref. \cite{al15}.  The two solutions can be distinguished more clearly
by their quadrupole moments,  314 and 341 fm$^2$ respectively.

\section{Appendix:  key functions in the codes}

The coded equations from the text above are listed here, together
with their location in the code.\\ 
Eq. (\ref{rho}):  {\tt util.calcRho}\\   
Eq. (\ref{vijkl}): {\tt hfsetup.mk\_vv}\\ 
Eq. (\ref{V}): {\tt util.calcV}\\ 
Eq. (\ref{Hsp},\ref{lamq}): {\tt util.calcHsp}\\ 
Eq. (\ref{EHF}): {\tt util.totalE}\\ 
Eq. (\ref{omega}): {\tt HFgradT}\\ 
Eq. (\ref{S}):  {\tt util2.entropy}\\ 
%Eq. (\ref{Z}): {\tt }\\ 
Eq. (\ref{Horb}):  {\tt util.calcOrbOp}\\ 
Eq. (\ref{hybrid},\ref{update}):  {\tt util2.updateU}\\ 
Eq. (\ref{reset},\ref{Qph},\ref{unitary}): {\tt util2.resetQ}\\ 
Eq. (\ref{Horb'}): {\tt util2.projectZ4}\\ 
Eq. (\ref{alpha'}):  {\tt util2.updatef}\\ 

%===========================================================  
\section{Acknowledgments}  
  
We would to thank Y. Alhassid and L. Robledo for discussions leading to this work, 
and H.
Nakada for the use of his Hartree-Fock code to validate the codes presented
here.  Support for this work was provided by the US Department of Energy
under Grant No. DE-FG02-00ER41132.

\end{document}